\documentclass[aps,prl,twocolumn,floatfix,superscriptaddress,preprintnumbers]{revtex4}
\usepackage{amsfonts}
\usepackage{amssymb}
\usepackage{amsmath,mathtools}

\RequirePackage{ifpdf}
\ifpdf
  \usepackage[pdftex]{graphicx}
\else
  \usepackage[dvipdfmx]{graphicx}
\fi

\newcommand{\braket}[1]{{\langle #1 \rangle}}

\begin{document}
\preprint{OU-HET-961}

\title{Deep Learning and AdS/CFT}
\author{Koji Hashimoto}
\author{Sotaro Sugishita}
\affiliation{Department of Physics, Osaka University, Toyonaka, Osaka 560-0043, Japan}
\author{Akinori Tanaka}
\affiliation{Mathematical Science Team, RIKEN Center for Advanced Intelligence Project (AIP),1-4-1 Nihonbashi, Chuo-ku, Tokyo 103-0027, Japan}
\affiliation{Department of Mathematics, Faculty of Science and Technology, Keio University, 3-14-1 Hiyoshi, Kouhoku-ku, Yokohama 223-8522, Japan}
\affiliation{interdisciplinary Theoretical \& Mathematical Sciences Program (iTHEMS) RIKEN 2-1, Hirosawa, Wako, Saitama 351-0198, Japan}
\author{Akio Tomiya}
\affiliation{Key Laboratory of Quark \& Lepton Physics (MOE) and Institute of Particle Physics,
Central China Normal University, Wuhan 430079, China}

\begin{abstract}
We present a deep neural network
representation of the AdS/CFT correspondence, and demonstrate the emergence of
the bulk metric function via the learning process for given data sets of response in boundary
quantum field theories.
The emergent radial direction of the bulk is identified with the depth of the layers,
and the network itself is interpreted as a bulk geometry. 
Our network provides a data-driven holographic modeling of strongly coupled systems.
With a scalar $\phi^4$ theory with unknown mass and coupling, 
in unknown curved spacetime with a black hole horizon, 
we demonstrate our deep learning (DL) framework can determine them
which fit given response data.
First, we show that, from boundary data generated by the AdS Schwarzschild spacetime, 
our network can reproduce the metric.
Second, we demonstrate that our network with experimental data as an input can determine
the bulk metric, the mass and the quadratic coupling of the holographic model.
As an example we use the experimental data of magnetic response of a strongly correlated material
Sm$_{0.6}$Sr$_{0.4}$MnO$_3$.
This AdS/DL correspondence not only enables gravity modeling of strongly correlated systems, but also
sheds light on a hidden mechanism of the emerging space in both AdS and DL.

\end{abstract}


\maketitle

\setcounter{footnote}{0}

\noindent
{\em Introduction.}---
The AdS/CFT correspondence
\cite{Maldacena:1997re,Gubser:1998bc,Witten:1998qj},
a renowned holographic relation between $d$-dimensional
quantum field theories (QFTs) and $(d+1)$-dimensional gravity, 
has been vastly applied to strongly coupled QFTs
including QCD and condensed matter systems. For phenomenology, 
the
holographic modelings 
were successful only for restricted class of systems in which
symmetries are manifest,
mainly because the mechanism of how the holography works is still
unknown. 
For a quantum system given,
we do not know whether its gravity dual exists and how we can construct 
a holographic model. 

Suppose one is given experimental data of linear/nonlinear response
under some external field, can one model it holographically? 
In this letter we employ {\em deep learning} (DL) \cite{Hinton,Bengio,LeCun}, 
an active subject of computational science,
to provide a {\rm data-driven} holographic gravity modeling of strongly coupled quantum systems.
While conventional holographic modeling starts with a given bulk gravity metric, our
novel DL method solves the inverse problem: given data of a boundary QFT calculates a suitable bulk metric
function, assuming the existence of a black hole horizon.

\begin{figure}
\includegraphics[width=6.5cm]{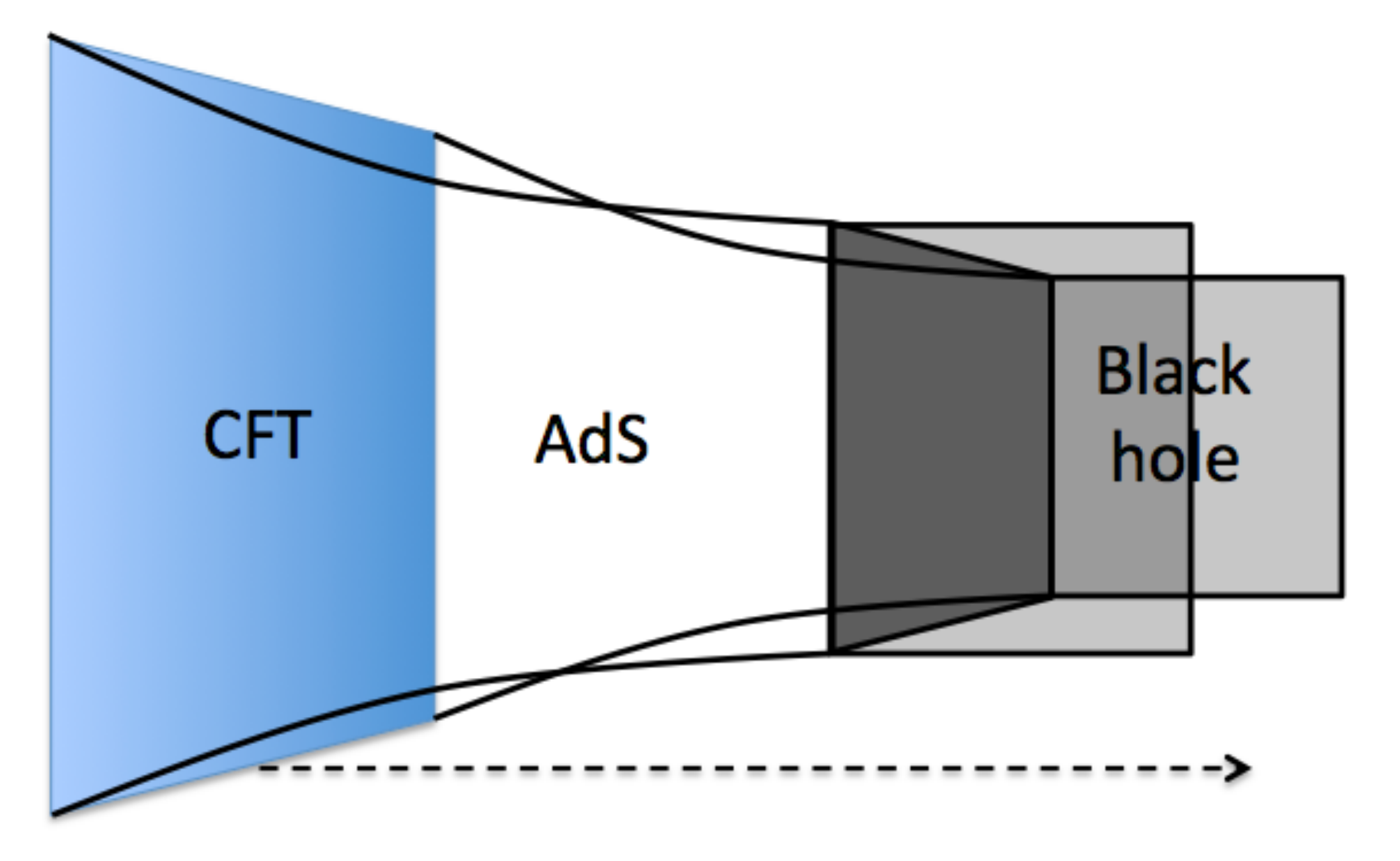}
\includegraphics[width=8.5cm]{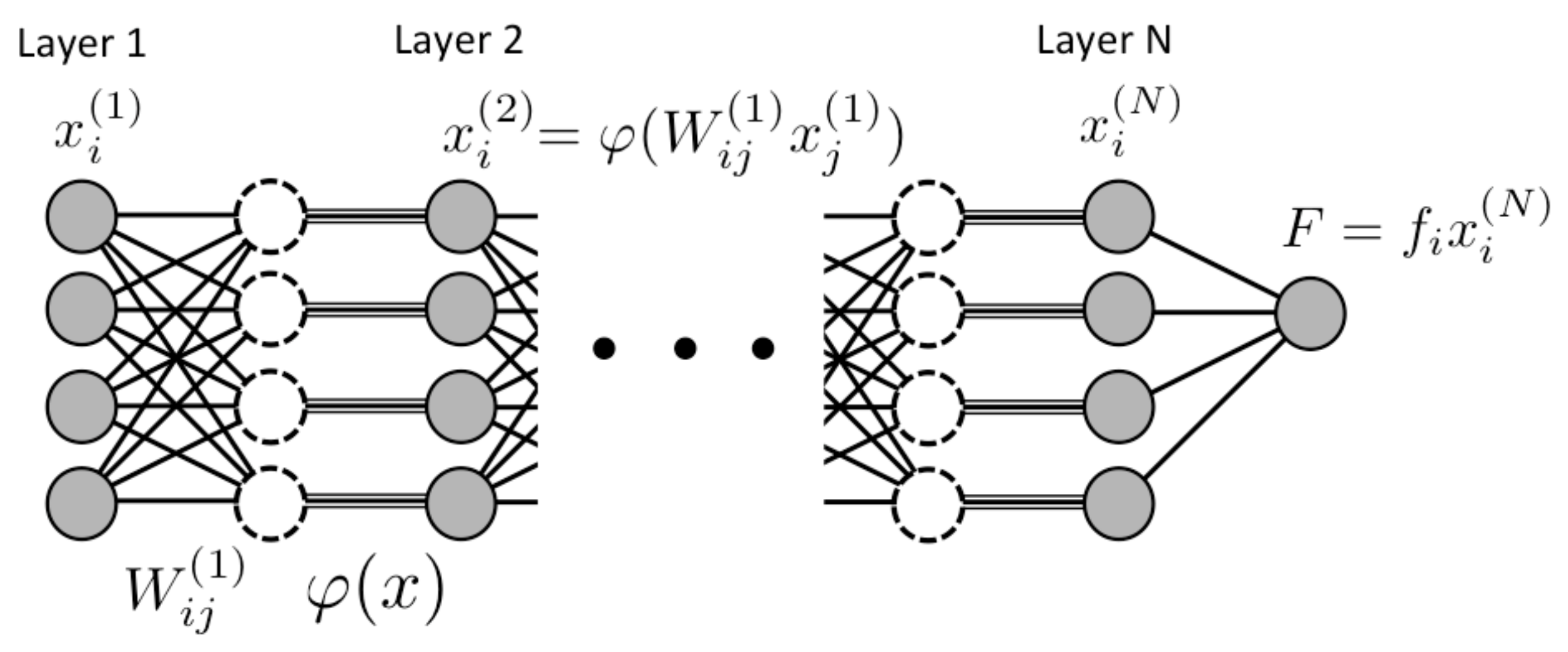}
\vspace*{-2mm}
\caption{The AdS/CFT and the DL. 
Top: a typical view of the AdS/CFT correspondence. The CFT at 
a finite temperature lives at a boundary of asymptotically AdS spacetime
with a black hole horizon at the other end.
Bottom: a typical neural network of a deep learning.
}
\label{fig:typ}
\end{figure}

Our strategy is simple: we provide a deep neural network representation
of a scalar field equation in $(d+1)$-dimensional curved spacetime. 
The discretized holographic (``AdS radial") direction is the deep layers, see Fig.~\ref{fig:typ}. 
The weights of the neural network
to be trained are identified with a metric component of the curved spacetime.
The input response data is at the boundary of AdS, and the output binomial data is the black hole horizon condition.
Therefore, a successful machine learning results in a concrete metric of a holographic modeling 
of the system measured by the experiment \footnote{We assume that the system can be described holographically by a classical scalar field in asymptotically AdS space.}.
We call this implementation of the holographic
model into the deep neural network as {\em AdS/DL correspondence}.

We check that the holographic DL modeling nicely works with the popular AdS Schwarzschild metric,
by showing that
the metric is successfully learned and reproduced by the DL. Then we proceed to use
an experimental data of a magnetic response of Sm$_{0.6}$Sr$_{0.4}$MnO$_3$
known to have strong quantum fluctuations, and demonstrate the emergence of a bulk metric via the
AdS/DL correspondence.

Our study gives a
first concrete implementation of the AdS/CFT into deep neural networks. We show the
emergence of a smooth geometry from given experimental data, which opens 
a possibility of revealing the mystery of the emergent geometry in
the AdS/CFT with the help of the active researches of DL.  
A similarity between the AdS/CFT and the DL was discussed recently \cite{You:2017guh} 
\footnote{See \cite{Gan:2017nyt,Lee:2017skk} for related essays. A continuum limit of the deep layers
was studied in a different context \cite{deepest}.}, and it can be discussed through
tensor networks, the AdS/MERA correspondence \cite{Swingle:2009bg} \footnote{
An application of DL or machine learning to quantum many-body problems
is a rapidly developing subject. See \cite{Science1} for one of the initial papers, together with 
recent papers \cite{dl1,dl2,dl3,dl4,dl5,dl6,dl7,dl8,dl9,dl10,dl11,dl12,dl13,dl14,dl15,dl16,dl17,dl18,dl19,dl20,dl21,dl21-2,dl22,dl23,dl24,dl25,dl26,dl27,dl28,dl29,dl30,dl31,dl32,dl33,dl34}.
For machine learning applied to string landscape, see 
\cite{He:2017aed,He:2017dia,Liu:2017dzi,Carifio:2017bov,Ruehle:2017mzq,Faraggi:2017cnh,Carifio:2017nyb}.}.

Let us briefly review 
a standard deep neural network. It  consists of layers (see Fig.~\ref{fig:typ}),  
and between the adjacent layers,
a linear transformation $x_i \to W_{ij} x_j$ and
a nonlinear transformation known as an activation function, $x_i \to \varphi(x_i)$,
are succeedingly act. The final layer is for summarizing all the component of the vector.
So the output of the neural network is 
\begin{align}
y(x^{(1)}) =  f_i \varphi(W_{ij}^{(N-1)}\varphi(W_{jk}^{(N-2)} \cdots \varphi(W_{lm}^{(1)}x_m^{(1)}))).
\label{nns}
\end{align}
In the learning process, the variables of the network $(f_i, W_{ij}^{(n)})$ for $n=1,2,\cdots, N-1$
are updated by a gradient descent method with a given loss function of the $L^1$-norm error,
\begin{align}
E \equiv \sum_{\rm data}\biggm| y(\bar{x}^{(1)})-\bar{y}\biggm| + E_{\rm reg}(W).
\label{loss}
\end{align}
Here the sum is over the whole set of pairs $\{(\bar{x}^{(1)},\bar{y})\}$ of the input data 
$\bar{x}^{(1)}$ and the output data $\bar{y}$. The regularization $E_{\rm reg}$ is introduced
to require expected properties for the weights \footnote{In Bayesian neural networks, regularizations are introduced
as a prior.}.

\vspace{5mm}
\noindent
{\em Neural network of scalar field in AdS.}---
Let us embed the scalar field theory into a deep neural network.
A scalar field theory in a $(d+1)$-dimensional curved spacetime is written as
\begin{align}
S\!=\!\int\!d^{d+1}x\sqrt{-\det g}\left[
-\frac12 (\partial_\mu \phi)^2-\frac12 m^2 \phi^2-V(\phi)
\right].
\label{scalar}
\end{align}
%
\begin{figure}
\includegraphics[width=8cm]{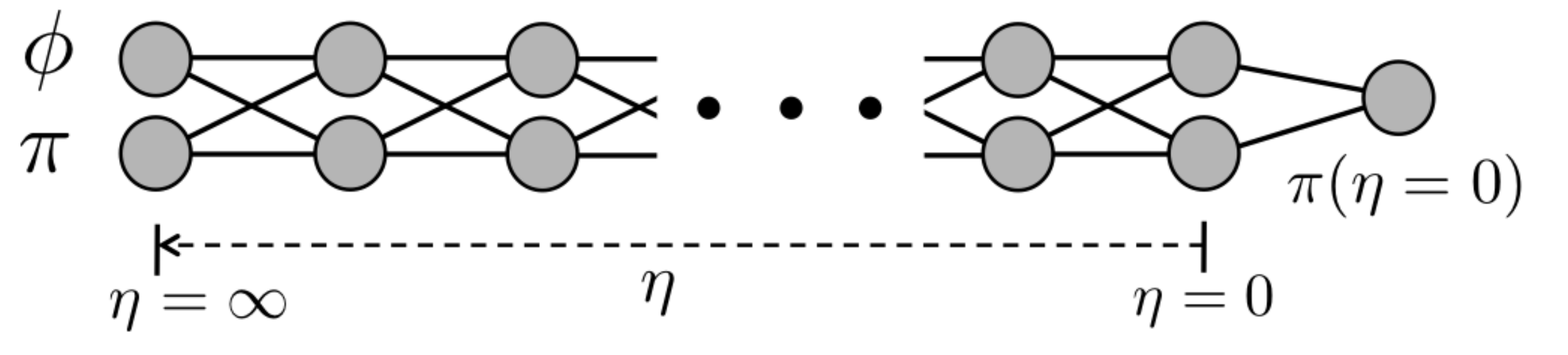}
\caption{
The simplest deep neural network reproducing the homogeneous scalar field equation
in a curved spacetime. Weights $W$ are shown by solid lines explicitly, while the activation is not.}
\label{fig:ournn}
\end{figure}
%
For simplicity we consider the field configuration to depend only on $\eta$
(the holographic direction).
Here the generic metric is given by
\begin{align}
ds^2=-f(\eta) dt^2 + d\eta^2 + g(\eta)(dx_1^2+\cdots +dx_{d-1}^2)
\label{genericm}
\end{align}
with the asymptotic AdS boundary condition $f \approx g \approx \exp[2\eta/L] \, (\eta \approx \infty)$
with the AdS radius $L$, and another boundary condition at the  black hole horizon,
$f \approx \eta^2, g \approx {}$const. $(\eta \approx 0)$. The classical equation of motion for $\phi(\eta)$
is
\begin{align}
\partial_\eta \pi + h(\eta)\pi -m^2\phi- \frac{\delta V[\phi]}{\delta \phi} = 0, \hspace{5mm} \pi \equiv \partial_\eta \phi \, ,
\label{sceq}
\end{align}
where we have defined $\pi$ so that the equations become a first order in derivatives. 
The metric dependence is summarized into a combination
$h(\eta)\equiv \partial_\eta \log\sqrt{f(\eta)g(\eta)^{d-1}}$.
Discretizing the radial $\eta$ direction, the equations are rewritten as
\begin{align}
&\phi(\eta+\Delta \eta) = \phi(\eta) + \Delta\eta \, \pi(\eta)\, , 
\label{prop}
\\
&\pi(\eta+\Delta \eta) = \pi(\eta) - \Delta\eta 
\left(h(\eta)\pi(\eta) -m^2\phi(\eta)- \frac{\delta V(\phi)}{\delta \phi(\eta)}\right).
\nonumber
\end{align}
We regard these equations as a propagation equation on a neural network, from the boundary $\eta=\infty$ 
where the input data $(\phi(\infty),\pi(\infty))$ is given, to the black hole horizon $\eta=0$ for
the output data, see Fig.~\ref{fig:ournn}. The $N$ layers of the deep neural network are a discretized
radial direction $\eta$ which is the emergent space in AdS, $\eta^{(n)}\equiv (N-n+1) \Delta \eta$.
The input data $x_i^{(1)}$ of the neural network is a two-dimensional real vector $(\phi(\infty),\pi(\infty))^{\rm T}$.
So the linear algebra part of the neural network (the solid lines in Fig.~\ref{fig:typ}) is automatically provided by 
\begin{align}
W^{(n)} = \left(
\begin{array}{cc}
1 & \Delta \eta \\
\Delta\eta \, m^2 & 1-\Delta \eta \,h(\eta^{(n)}) 
\end{array}
\right).
\label{W}
\end{align}
The activation function at each layer reproducing \eqref{prop}
is 
\begin{align}
\left\{
\begin{array}{l}
\varphi(x_1) = x_1,
\\ \varphi(x_2) = x_2 + \Delta \eta \, \frac{\delta V(x_1)}{\delta x_1} \, .
\end{array}
\right.
\label{act}
\end{align}
The definitions
\eqref{W} and \eqref{act} bring 
the scalar field system in curved geometry \eqref{scalar} into 
the form of the neural network \eqref{nns} \footnote{
Note that $\varphi(x_2)$ in \eqref{act} includes $x_1$ so it is not local,
opposed to the standard neural network \eqref{nns}
with local activation functions. See the supplemental material for 
an improved expression with local activation functions.
}.

\begin{figure}
\vspace{-5mm}
\includegraphics[width=8.5cm]{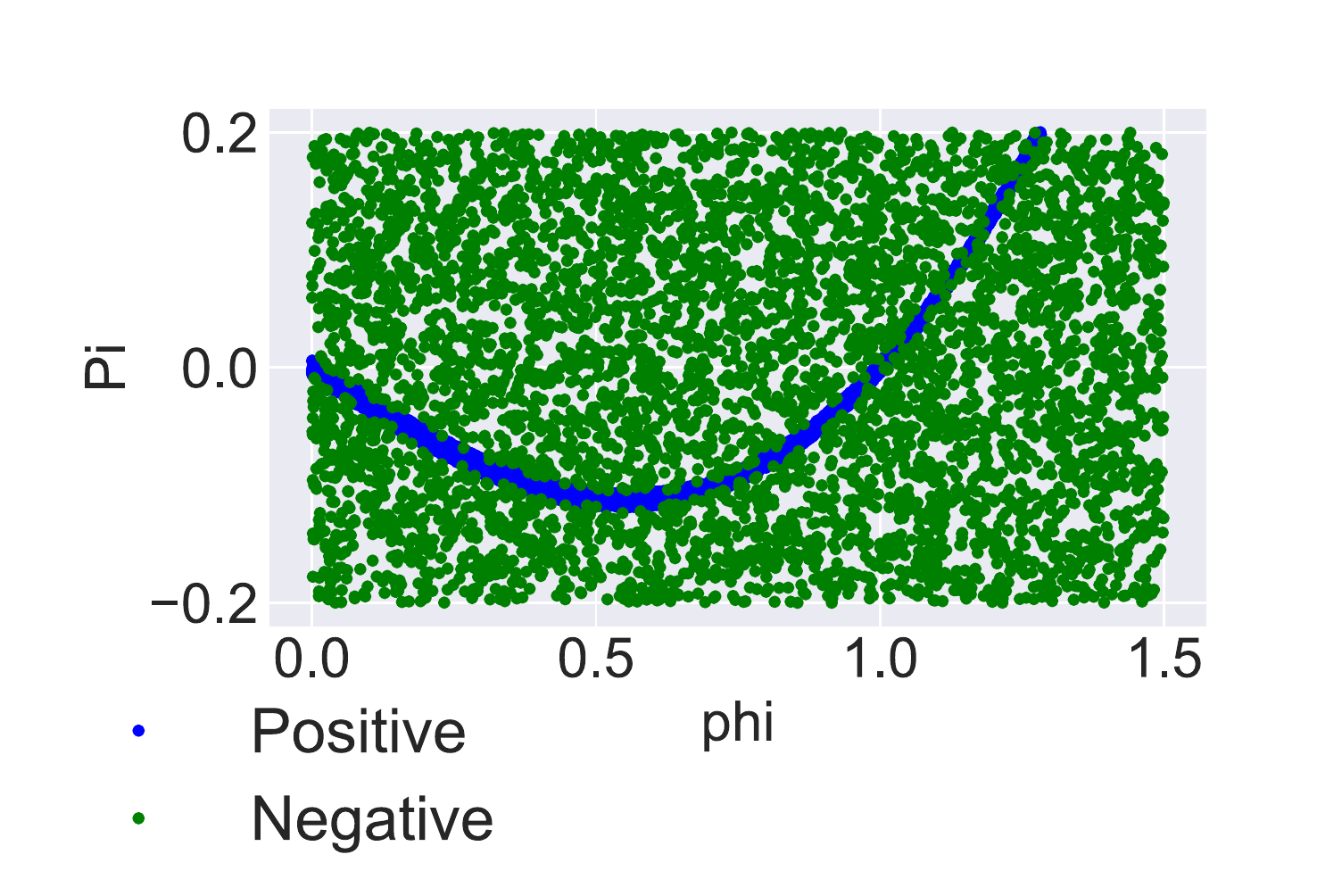}
\caption{
The data generated by the discretized AdS Schwarzschild metric \eqref{AdSS}. Blue points are the positive 
data $(y=0)$
and the green points are the negative data $(y=1)$.}
\label{fig:data}
\end{figure}

\begin{figure*}
\vspace{-5mm}
\includegraphics[width=18cm]{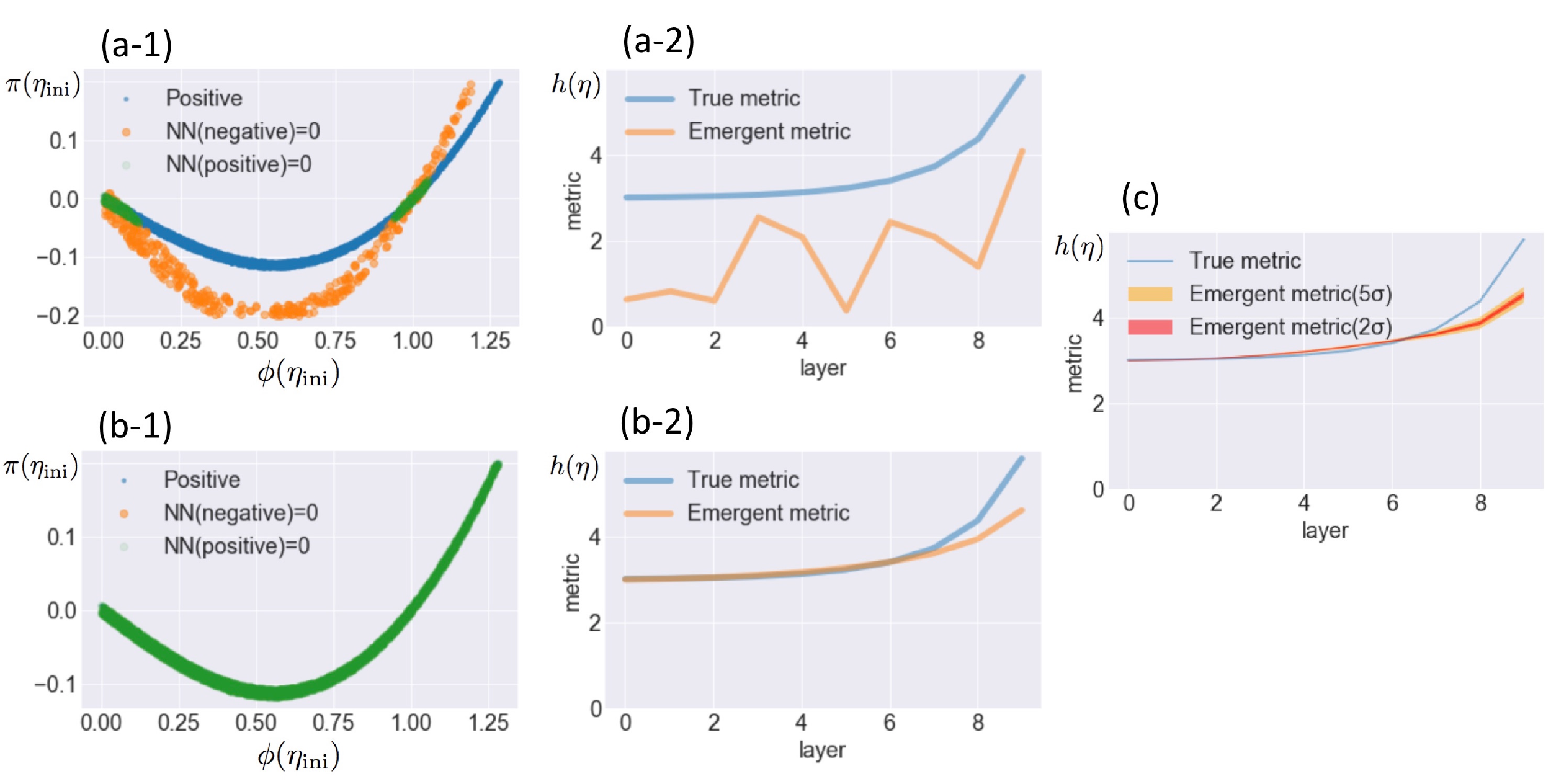}
\caption{
Before the learning (a) and after the learning (b).
(a-1) The $(\phi,\pi)$ plot at the first epoch. Blue and green dots are positive data. 
Orange and green dots are data judged as
``positive" by using the initial trial metric. (a-2) 
The orange line is the initial trial metric (randomly generated), while the blue line is
the discretized AdS Schwarzschild metric \eqref{AdSS}.
(b-1)  The $(\phi,\pi)$ plot after the training for 100 epochs.
(b-2) The learned metric (orange line) almost coincides with the original AdS Schwarzschild metric,
which means our neural network successfully learned the bulk metric.
(c) Statistical analysis of 50 learned metrics.
}
\label{fig:adslearn}
\end{figure*}

\vspace{5mm}
\noindent
{\em Response and input/output data.}---
In the AdS/CFT, asymptotically AdS spacetime provides a boundary condition of
the scalar field corresponding to the response data of the quantum field theory (QFT). 
With the AdS radius $L$, asymptotically $h(\eta)\approx d/L$. 
The external field value
$J$ (the coefficient of a non-normalizable mode of $\phi$) 
and its response $\braket{\cal O}$ (that of a normalizable mode) in the QFT are \cite{Klebanov:1999tb},
in the unit of $L=1$, 
a linear map
\begin{align}
&\phi(\eta_{\rm ini}) = J \exp[-\Delta_- \eta_{\rm ini}] + \braket{\cal O}\frac{\exp[-\Delta_+ \eta_{\rm ini}]}{\Delta_+-\Delta_-},
\label{pia}
\\
&\pi(\eta_{\rm ini}) = -J \Delta_-\exp[-\Delta_- \eta_{\rm ini}] 
-\braket{\cal O}\frac{\Delta_+\exp[-\Delta_+ \eta_{\rm ini}]}{\Delta_+-\Delta_-},
\nonumber
\end{align}
with $\Delta_\pm \equiv (d/2)\pm\sqrt{d^2/4 + m^2L^2}$ ($\Delta_+$ 
is the conformal dimension of the QFT operator ${\cal O}$
corresponding to the bulk scalar $\phi$). The value $\eta=\eta_{\rm ini} \approx\infty$ is the 
regularized cutoff of the asymptotic AdS spacetime. We use \eqref{pia} for converting the response data
of QFT to the input data of the neural network.

The input data at $\eta=\eta_{\rm ini}$ propagates in the neural network toward $\eta=0$, the horizon. 
If the input data is positive, the output at the final layer should satisfy the boundary condition of the black hole
horizon (see for example \cite{Horowitz:2010gk}),
\begin{align}
0=F\equiv \left[\frac{2}{\eta} \pi -m^2 \phi - \frac{\delta V(\phi)}{\delta \phi}\right]_{\eta=\eta_{\rm fin}}
\label{Ff}
\end{align}
Here $\eta=\eta_{\rm fin} \approx 0$ is the horizon cutoff. Our final layer is defined by
the map $F$, and the output data is $y=0$ for a positive answer response data $(J,\braket{\cal O})$.
In the limit $\eta_{\rm fin}\to 0$, the condition \eqref{Ff} is equivalent to $\pi(\eta=0)=0$.

With this definition of the network and the training data, we can make the  deep neural network to
learn the metric component function $h(\eta)$, the parameter $m$ and the interaction $V[\phi]$.
The training is with a loss function $E$ given by \eqref{loss}\footnote{The explicit expression
for the loss function is available for $\lambda=0$: see the supplemental material.}. Experiments provide only
positive answer data $\{(J,\braket{\cal O}),y=0\}$, 
while for the training we need also negative answer data : $\{(J,\braket{\cal O}),y=1\}$.
It is easy to generate false response data $(J,\braket{\cal O})$, and we assign output $y=1$ for them. 
To make the final output of the neural network to be binary,
we use a function $\tanh |F|$ (or its variant) for the final layer rather than just $F$, 
because $\tanh |F|$
provides $\approx 1$ for any negative input.

\begin{figure}[b]
\includegraphics[width=3cm]{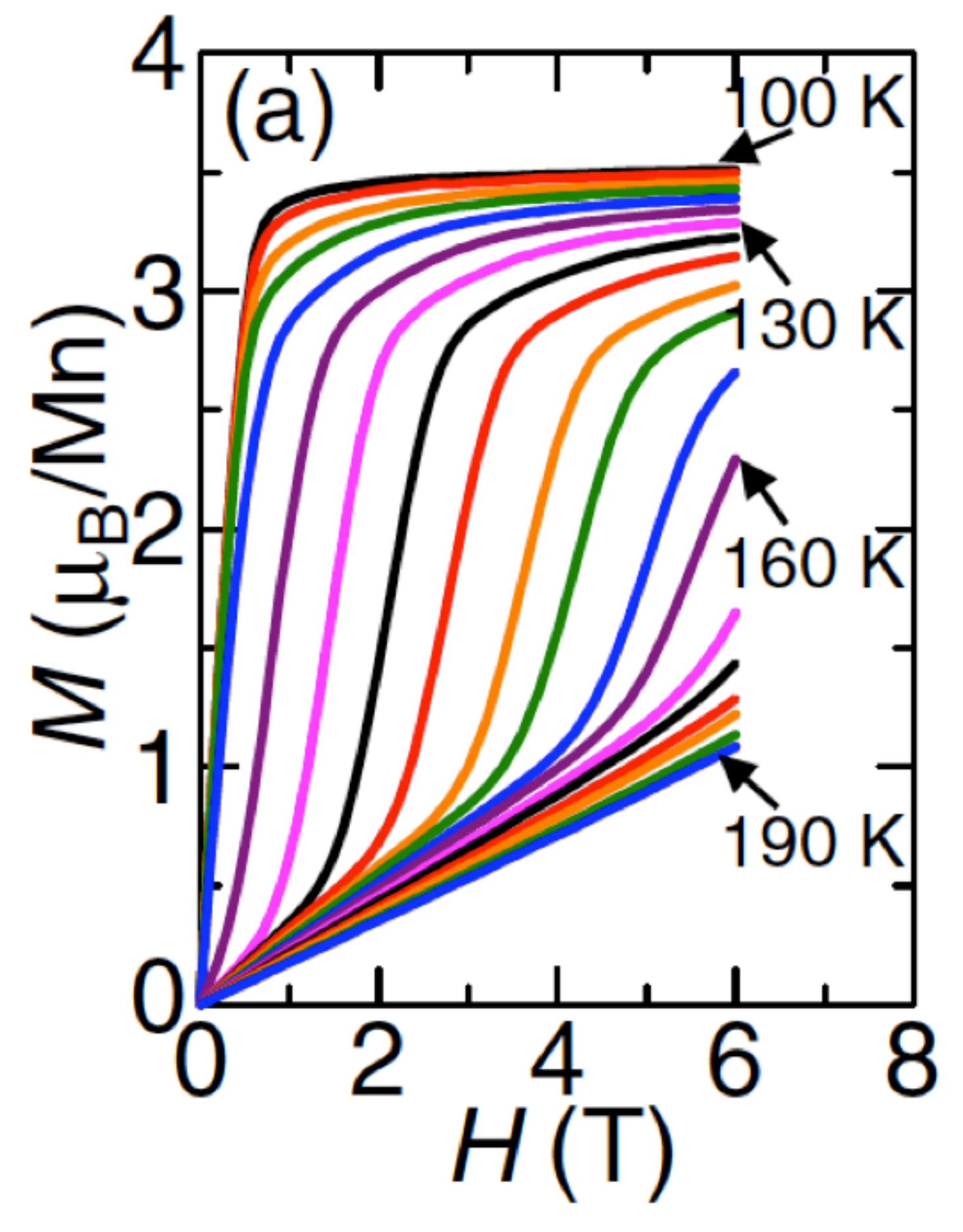}
\includegraphics[width=5.5cm]{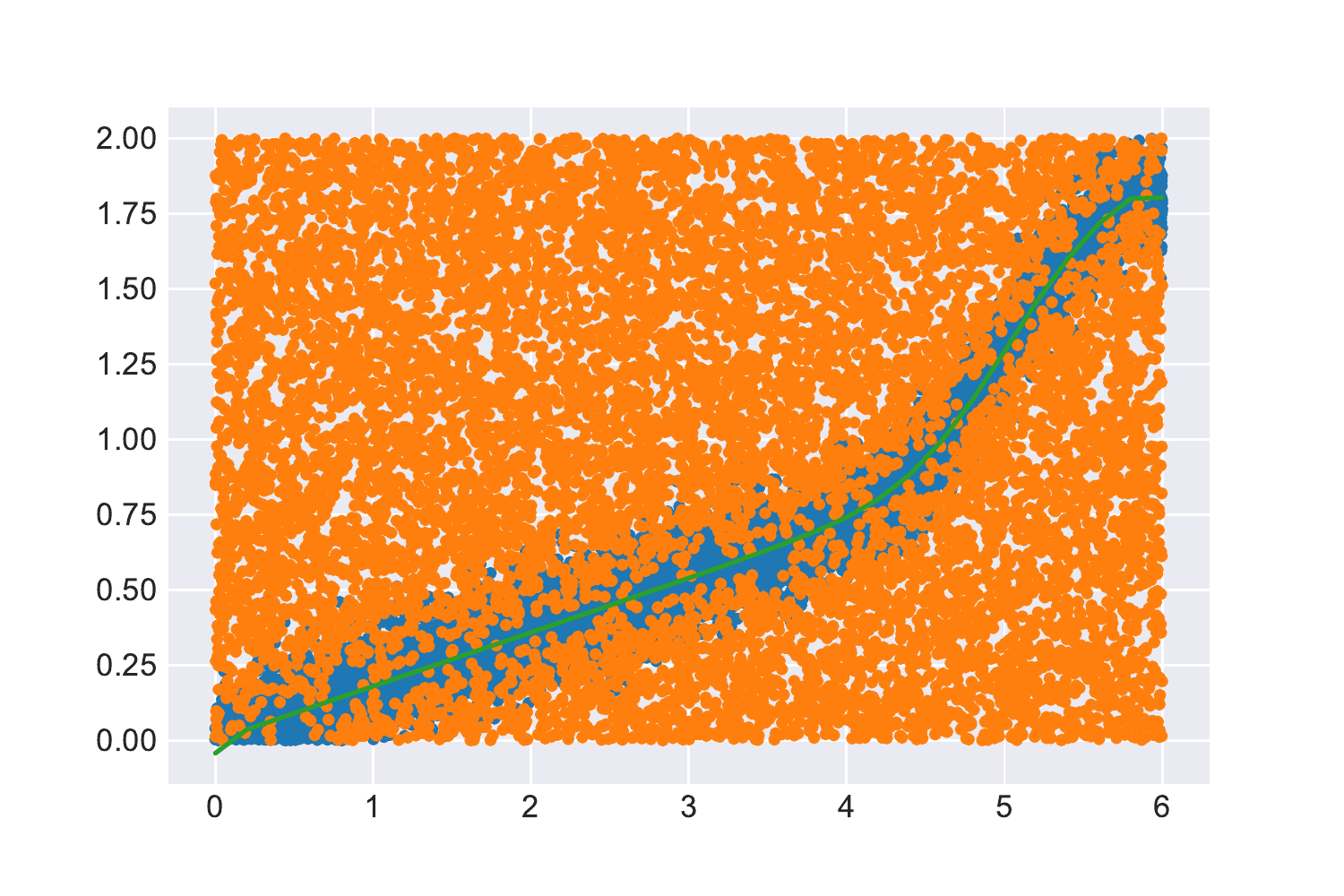}
\caption{Left: Experimental data of magnetization ($M$) versus magnetic field ($H$) for the material
Sm$_{0.6}$Sr$_{0.4}$MnO$_3$. Figure taken from \cite{sakai}. Right: Positive (blue) and negative
(orange)
data sets generated by the experimental data at the temperature $155$ K, with a random noise added.}
\label{fig:sakai}
\end{figure}

\begin{figure*}
\includegraphics[width=18cm]{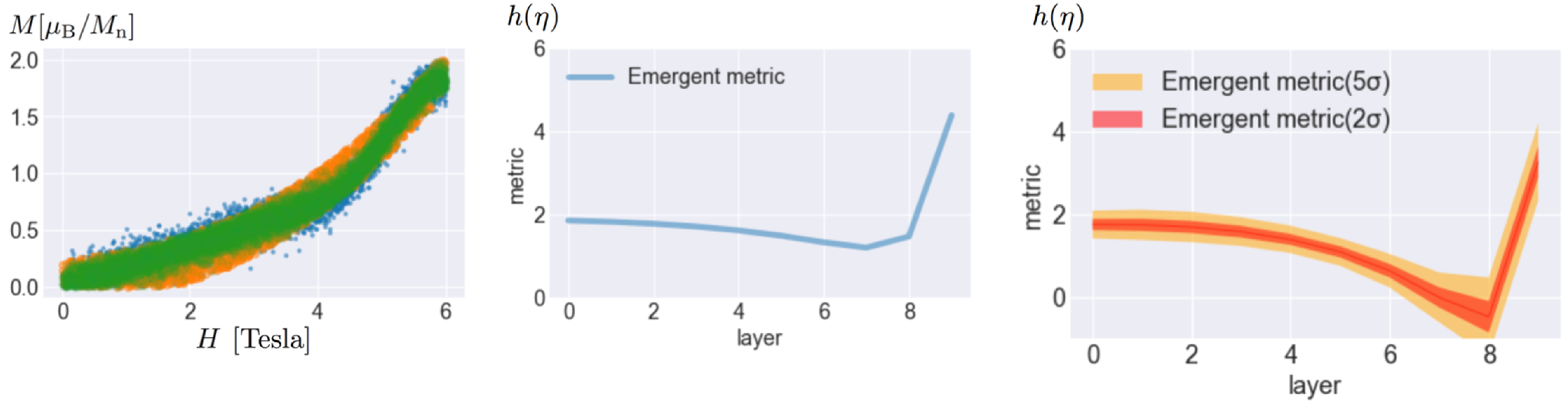}
\caption{Left: A result of the machine learning for 
fitting of the experimental data. Blue and green dots are positive experimental data. 
Orange and green dots are data judged as
``positive" by using the learned metric (Center). The total loss after the training is 0.0096. 
Right: Statistical average of the 13 learned metrics all of whose loss are less than 0.02.}
\label{fig:expres}
\end{figure*}

\vspace{5mm}
\noindent
{\em Learning test: AdS Schwarzschild black hole.}---
To check whether this neural network can learn the bulk metric, we first demonstrate a learning test.
We will see that with data generated by a known AdS Schwarzschild metric, our neural network can learn
and reproduce the metric\footnote{See the supplemental material for the details
about the coordinate system.}.
We work here with $d=3$ in the unit $L=1$. The metric is
\begin{align}
h(\eta)= 3\coth(3\eta)
\label{AdSS}
\end{align}
and we discretize the $\eta$ direction by $N=10$ layers with $\eta_{\rm ini}=1$ and $\eta_{\rm fin}=0.1$.
We fix for simplicity $m^2=-1$ and $V[\phi]=\frac{\lambda}{4} \phi^4$ with $\lambda = 1$.
Then we generate positive answer data with the neural network with the discretized \eqref{AdSS},
by collecting randomly generated $(\phi(\eta_{\rm ini},\pi(\eta_{\rm ini}))$ giving $|F|<\epsilon$
where $\epsilon=0.1$ is a cut-off.
The negative answer data are similarly generated under the criterion $|F|>\epsilon$. We collect
1000 positive and 1000 negative data, see Fig.~\ref{fig:data}.
Since we are interested in a smooth continuum limit of $h(\eta)$, and the horizon boundary condition $h(\eta)\approx 1/\eta (\eta \approx 0)$, we introduced the regularization 
$E_{\rm reg}^{(1)} \equiv c_{\rm reg}\sum_{n=1}^{N-1}(\eta^{(n)})^4 (h(\eta^{(n+1)})-h(\eta^{(n)}))^2 \propto \int d\eta\, (h'(\eta)\eta^2)^2$, with $c_{\rm reg}=10^{-3}$.

We use PyTorch for a Python deep learning library to implement our network
\footnote{See the supplemental material for the details of the setup and coding,
and the effect of the regularization and statistics.}.
The initial metric is randomly chosen. Choosing the batch size equal to 10,
we find that after 100 epochs of the training our deep 
neural network successfully learned $h(\eta)$ and it coincides with
\eqref{AdSS}, see Fig.~\ref{fig:adslearn} (b)
\footnote{At the first epoch, the loss was 0.2349, while after the 100th epoch, the loss was 0.0002. 
We terminated the learning when the loss did not decrease.}.
The statistical analysis with 50 learned metric, Fig.~\ref{fig:adslearn} (c), shows that
the asymptotic AdS region is almost perfectly learned.  The near horizon region has $\approx 30 \%$ 
systematic error,  and it is expected also for the following analysis with experimental data.

\vspace{5mm}
\noindent
{\em Emergent metric from experiments.}---
Since we have checked that the AdS Schwarzschild metric is successfully reproduced, we shall apply
the deep neural network to learn a bulk geometry for a given experimental data. We use 
experimental data of the magnetization curve (the magnetization $M [\mu_{\rm B}/M_{\rm n}]$ 
vs the external magnetic field $H$ [Tesla])
for the 3-dimensional material Sm$_{0.6}$Sr$_{0.4}$MnO$_3$ which is known to
have a strong quantum fluctuation \cite{sakai}, see Fig.~\ref{fig:sakai}.
We employ a set of data at temperature $155$ K which is slightly above the critical temperature, 
since it exhibits the deviation from a linear $M$-$H$ curve
suggesting a strong correlation.
To form a positive data we add a random noise
around the experimental data, and also generated negative data positioned away from the positive data.\footnote{
Our experimental data does not have an error bar, so we add the noise.}

The same neural network is used, except that we add a new zero-th layer
to relate the experimental data with $(\phi, \pi)$, motivated by \eqref{pia} : 
\begin{align}
\left. \begin{array}{ll}
\phi(\eta_\text{ini})
=
\alpha H + \beta M
\\
\pi(\eta_\text{ini})
=
- \Delta_- \alpha H
- \Delta_+ \beta M.
\end{array} \right.
\label{nnpia2}
\end{align}
We introduce the normalization parameters $\alpha$ and $\beta$ to relate $(H,M)$ to the bulk $\phi$,
and the asymptotic AdS radius $d/h(\infty)\equiv L$ is included in
$\Delta_\pm = (d/2)\left(1\pm\sqrt{1 + 4m^2/h(\infty)^2}\right)$.
In our numerical code we introduce a dimensionful parameter $L_{\rm unit}$
with which all the parameters are measured in the unit $L_{\rm unit}=1$.
We add another regularization term $E_{\rm reg}=E_{\rm reg}^{(1)} + E_{\rm reg}^{(2)}$
with $E_{\rm reg}^{(2)}\equiv c_{\rm reg}^{(2)} (h(\eta^{(N)})-1/\eta^{(N)})^2$
which forces $h(\eta^{(N)})$, the metric value near the horizon, to match
the standard horizon behavior $1/\eta$, see the supplemental material for the details. 
We chose $N=10$ and $c_{\rm reg}^{(2)}=10^{-4}$.
In the machine learning, $m$ and $\lambda$, $\alpha$ and $\beta$ are trained, as well as the metric function
$h(\eta)$. 

We stopped the training when the loss becomes smaller than 0.02, and collected 13 successful cases.
The emergent metric function $h(\eta)$ obtained by the machine learning is shown in Fig.~\ref{fig:expres}. 
It approaches a constant at the boundary, meaning that it is properly an asymptotically AdS spacetime. 
The obtained (dimensionless) parameters for the scalar field are $m^2 L^2= 5.6 \pm 2.5$, 
$\lambda/L = 0.61 \pm  0.22$ \footnote{For numerically estimated conformal dimension and
its implications, see the supplemental material.}.
In this manner, a holographic model is determined numerically from the experimental data, by the DL.

\vspace{5mm}
\noindent
{\em Summary and outlook.}--- 
We put a bridge between two major subjects about hidden dimensions: the AdS/CFT and the DL.
We initiate a data-driven holographic modeling of quantum systems
by formulating the gravity dual on a deep neural network. We show that with an appropriate choice of
the sparse network and the input/output data the AdS/DL correspondence is properly formulated, 
and the standard
machine learning works nicely for the automatic emergence of the bulk gravity for given response
data of the boundary quantum systems.

Our method can be applied to any holographic models. With vector fields in the bulk, not only $h(\eta)$
but other metric components can be determined by the DL. 
To explore the significance of the neural network representation of black hole horizons, 
the systematic error near the horizon would need to be reduced. Comparison with confining gauge theories 
giving a Dirichlet condition as the output could be helpful.

How can our study shed light on the mystery of the emergent spacetime in AdS/CFT correspondence?
A continuum limit of deep neural networks can accommodate arbitrarily nonlocal systems
as the network basically includes all-to-all inter-layer connections. 
So, the emergence of the new spatial dimension would need a reduction of the full DL parameter space.
A criterion to find a properly sparse neural network
which can accommodate local bulk theories is missing, and the question is similar to the AdS/CFT 
where criteria for QFT to have a gravity dual is still missing.
At the same time, 
our work suggests that the bulk emergence could be a more generic phenomenon.
For further exploration of the AdS/DL correspondence, we plan to formulate
a ``holographic autoencoder", motivated by a similarity between DL autoencoders and 
the cMERA at finite temperature 
\cite{Matsueda:2012xm,Mollabashi:2013lya}, 
and also the thermofield formulation of the AdS/CFT \cite{Maldacena:2001kr,Hartman:2013qma}.
Characterization of black hole horizons in DL may be a key to understand the bulk emergence.

\vspace{5mm}
\begin{acknowledgments}
We would like to thank H.~Sakai for providing us with the experimental data.
K.~H.~would like to thank S.~Amari, T.~Ohtsuki and N.~Tanahashi 
for valuable discussions.
The work of K.~H.~was supported 
in part by JSPS KAKENHI Grants No.~JP15H03658, 
No.~JP15K13483, and No.~JP17H06462. 
S.~S.~is supported in part by the Grant-in-Aid for JSPS Research Fellow, Grant No.~JP16J01004.
The work of A.~Tanaka was supported by the RIKEN Center for AIP.
A.~Tomiya was fully supported by Heng-Tong Ding.
The work of A.~Toimya was supported in part by NSFC under grant no. 11535012.
\end{acknowledgments}


\clearpage 


\renewcommand{\theequation}{S.\arabic{equation}}

\begin{widetext}
\section*{\large Supplemental Material for ``Deep Learning and AdS/CFT"}

\section{Hamiltonian systems realized by deep neural network} 

Here we show that a restricted class of Hamiltonian systems can be realized by a deep 
neural network with a local activation function.\footnote{Here we regard the time evolution of the 
Hamiltonian as the propagation in the neural network. For other ways to identify Hamiltonian systems
in machine learning, see \cite{Tegmark}.}
We consider a generic Hamiltonian $H(p,q)$ and its Hamilton equation, and seek for 
a deep neural network representation \eqref{nns} representing the time evolution by $H(p,q)$.
The time direction is discretized to form the layers. (For our AdS/CFT examples, the radial
evolution corresponds to the time direction of the Hamiltonian which we consider here.) 

Let us try first the following generic neural network and identify the time translation $t\to t + \Delta t$
with the inter-layer propagation,
\begin{align}
 q(t+\Delta t) = \varphi_1(W_{11}q(t)+W_{12}p(t)) ,  \quad 
 p(t+\Delta t) = \varphi_2(W_{12}q(t)+W_{22}p(t)) .
 \label{geneH}
\end{align}
This is successive actions of a linear $W$ transformation and a local $\varphi$ nonlinear transformation.
The relevant part of the network is shown in the left panel of Fig.~\ref{fig:ham}.
The units $x_1^{(n)}$ and $x_2^{(n)}$ are directly identified with the canonical variables $q(t)$ and $p(t)$,
and $t = n \Delta t$.
We want to represent Hamilton equations to be of the form \eqref{geneH}.
It turns out that it is impossible except for free Hamiltonians.

In order for \eqref{geneH} to be consistent at $\Delta t=0$, we need to require
\begin{align}
W_{11} = 1 + {\cal O}(\Delta t), \quad
W_{22} = 1 + {\cal O}(\Delta t), \quad
W_{12} =  {\cal O}(\Delta t), \quad
W_{21} =  {\cal O}(\Delta t), \quad
\varphi(x) = x + {\cal O}(\Delta t).
\end{align}
So we put an ansatz
\begin{align}
W_{ij} = \delta_{ij} + w_{ij} \Delta t, \quad
\varphi_i(x) = x + g_i(x) \Delta t,
\end{align}
where $w_{ij}$ $(i,j=1,2)$ are constant parameters and $g_i(x)$ $(i=1,2)$ are nonlinear functions. Substituting these
into the original \eqref{geneH} and taking the limit $\Delta t \to 0$, we obtain
\begin{align}
\dot{q}=w_{11}q+w_{12}p+g_1(q), \quad \dot{p}=w_{21}q+w_{22}p+g_2(p) \, .
\end{align}
For these equations to be Hamiltonian equations, we need to require a symplectic structure
\begin{align}
\frac{\partial}{\partial q}\left(w_{11}q+w_{12}p+g_1(q)\right) 
+\frac{\partial}{\partial p}\left(w_{21}q+w_{22}p+g_2(p)\right)=0 .
\end{align}
However, this equation does not allow any nonlinear activation function $g_i(x)$. 
So, we conclude that a simple identification of the units of the neural network 
with the canonical variables allow only linear
Hamilton equations, thus free Hamiltonians.

In order for a deep neural network representation to allow generic nonlinear Hamilton equations, 
we need to improve our identification of the units with the canonical variables, and also of
the layer propagation with the time translation.
Let us instead try
\begin{align}
x_i(t+\Delta t) = \widetilde{W}_{ij}\varphi_j(W_{jk}x_k(t)) .
\end{align}
The difference from \eqref{geneH} is two folds: First, we define $i,j,k=0,1,2,3$ with $x_1=q$ and $x_2=p$,
meaning that we have additional units $x_0$ and $x_3$. Second, we consider a multiplication by a linear
$\widetilde{W}$. So, in total, 
this is successive actions of a linear $W$, a nonlinear local $\varphi$ and a linear $\widetilde{W}$,
and we interpret this set as a time translation $\Delta t$. 
Since we pile up these sets as many layers, 
the last $\widetilde{W}$ at $t$ and the next $W$ at $t+\Delta t$
are combined into a single linear transformation $W_{t+\Delta t} \widetilde{W}_t$, so the standard form \eqref{nns}
of the deep neural network is kept.

We arrange the following sparse weights and local activation functions
\begin{align}
W = \left(
\begin{array}{cccc}
0 &0&v&0 \\
0 & 1 + w_{11}\Delta t & w_{12}\Delta t & 0 \\
0 & w_{21}\Delta t & 1+w_{22}\Delta t & 0 \\
0 &u&0&0 
\end{array}
\right), \quad 
\widetilde{W} = \left(
\begin{array}{cccc}
0 &0&0&0 \\
\lambda_1 & 1 & 0 & 0 \\ 
0 & 0 & 1 & \lambda_2 \\
0 &0&0&0 
\end{array}
\right), \quad 
 \left(
\begin{array}{l}
\varphi_0(x_0)\\
\varphi_1(x_1)\\
\varphi_2(x_2)\\
\varphi_3(x_3)
\end{array}
\right)
= \left(
\begin{array}{c}
f(x_0)\Delta t
\\
1
\\
1
\\
g(x_3)\Delta t
\end{array}
\right), 
\end{align}
where $u,v, w_{ij}$ ($i,j=1,2$) are constant weights, and $\varphi_i(x_i)$ are local activation functions.
The network is shown in the right panel of Fig.~\ref{fig:ham}.
Using this definition of the time translation, we arrive at
\begin{align}
\dot{q}=w_{11}q+w_{12}p+\lambda_1 f(vp), \quad
\dot{p}=w_{11}q+w_{12}p+\lambda_2 g(uq).
\end{align}
Then the symplectic constraint means $w_{11}+w_{22}=0$, and the Hamiltonian is given by
\begin{align}
H = w_{11}pq + \frac12 w_{12}p^2 -\frac12 w_{21}q^2+\frac{\lambda_1}{v} F(vp) -\frac{\lambda_2}{u}  G(uq)
\label{Hg}
\end{align} where
$F'(x_0)=f(x_0)$ and $G'(x_3)=g(x_3)$. This is the generic form of the nonlinear Hamiltonians which admit
a deep neural network representation. Our scalar field equation in the curved geometry \eqref{sceq} is within this
category.

For example, choosing 
\begin{align}
w_{11}=w_{21}=0, \quad w_{12}=1/m, \quad \lambda_1=0, \quad \lambda_2=1, \quad u=1,
\end{align}
means a popular Hamiltonian for a non-relativistic particle moving in a potential,
\begin{align}
H = \frac1{2m} p^2 - G(q).
\end{align}
A more involved identification of the time translation and the layer propagation may be able to accommodate
Hamiltonians which are not of the form \eqref{Hg}. We leave generic argument for the future investigation.

\begin{figure}
\includegraphics[width=6cm]{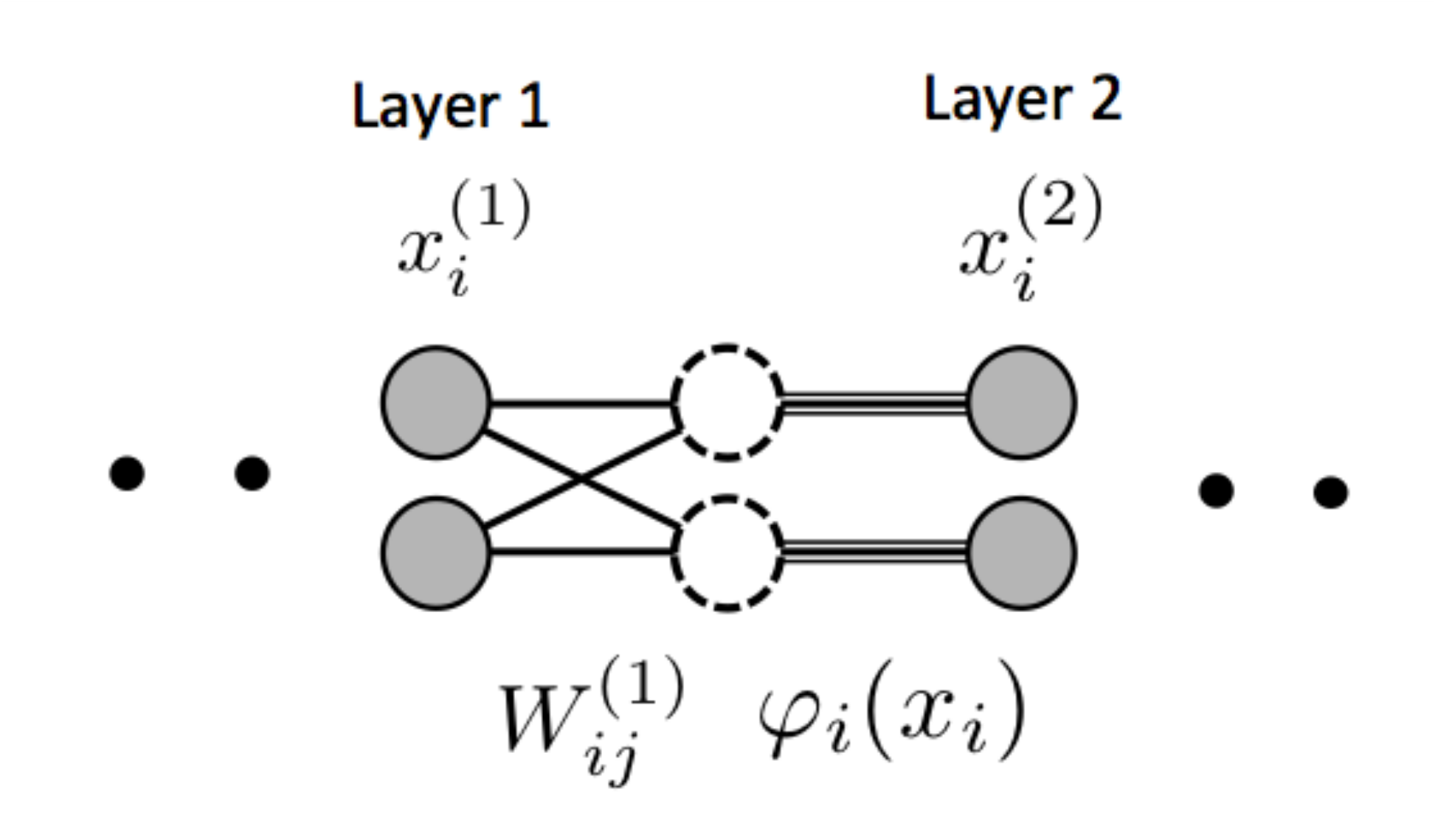}
\hspace{10mm}
\includegraphics[width=8cm]{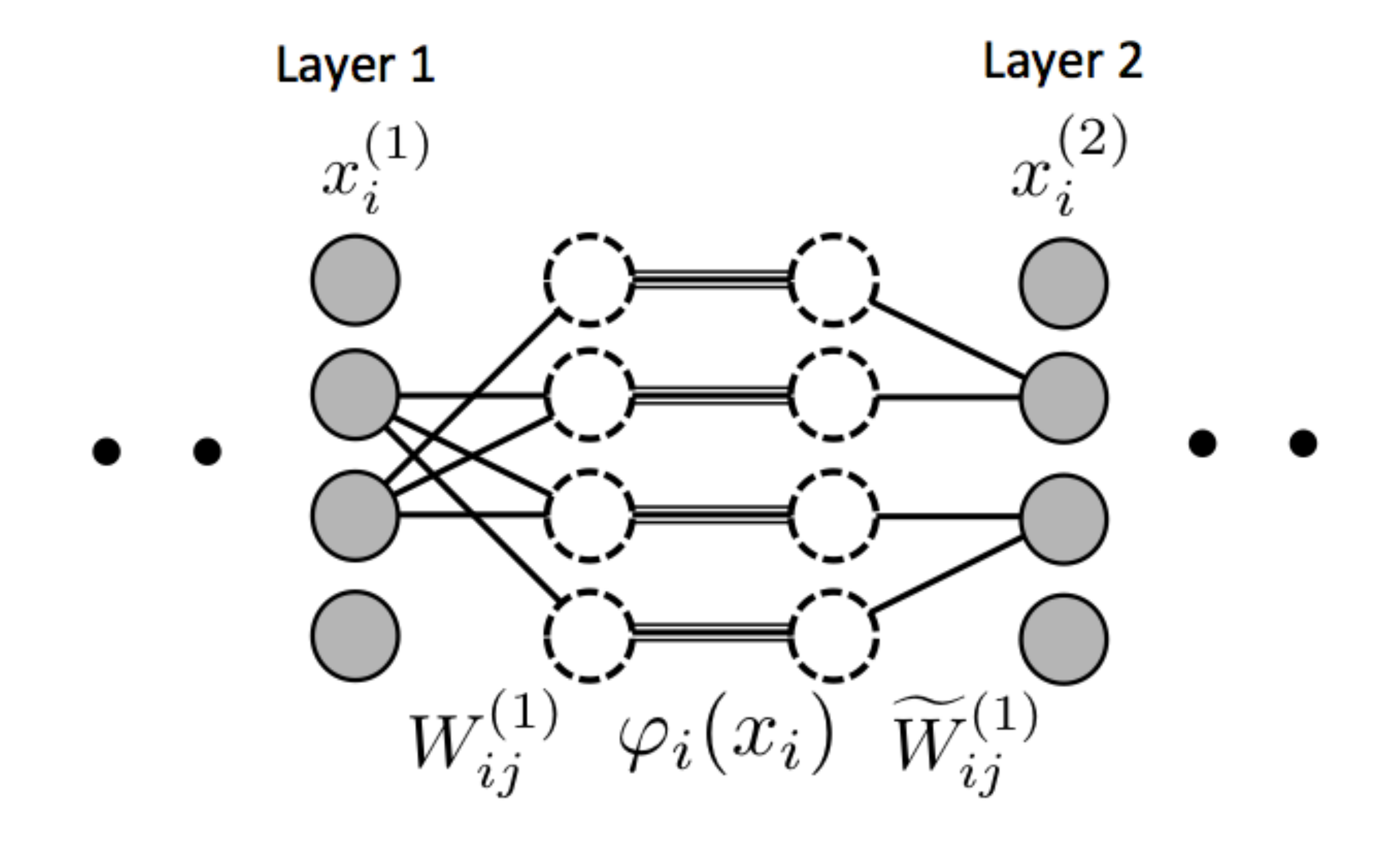}
\caption{
Left: a naive identification of the canonical variables $q,p$ and the units, and of
the time translation with the inter-layer propagation. Right: an improved neural network
whose continuum limit provides a nonlinear Hamilton system.}
\label{fig:ham}
\end{figure}

\section{Error function of the AdS scalar system}

For $\lambda=0$, we can obtain an explicit expression for the error function (loss function) for the 
machine learning in our AdS scalar field system.
The scalar field equation \eqref{sceq} can be formally solved as a path-ordered form
\begin{align}
\left(
\begin{array}{c}
\pi(\eta)\\
\phi(\eta)
\end{array}
\right)
= {\rm P}\exp\left\{
\int_\eta^{\eta_{\rm ini}}\! d\tilde{\eta}
\left(
\begin{array}{cc}
h(\tilde{\eta}) & -m^2\\
-1 & 0
\end{array}
\right)
\right\}
\left(
\begin{array}{c}
\pi(\eta_{\rm ini})\\
\phi(\eta_{\rm ini})
\end{array}
\right).
\end{align}
So, in the continuum limit of the discretized neural network,
the output is provided as
\begin{align}
\tanh |\pi(0)|
=\tanh
\left[
(1 \; 0)\,
 {\rm P}\!\exp\left\{
 \int_0^\infty \!d\tilde{\eta}
\left(
\begin{array}{cc}
h(\tilde{\eta}) & -m^2\\
-1 & 0
\end{array}
\right)
\right\}
\left(
\begin{array}{c}
\pi(\infty)\\
\phi(\infty)
\end{array}
\right)
\right]
\end{align}
Then the error function \eqref{loss} is provided as
\begin{align}
E[h(\eta)] = &
\sum_{\small\begin{array}{l}
\{\pi(\infty),\phi(\infty)\}
\\{\rm positive }
\end{array}}
\left(
\tanh
\left[
(1 \; 0)\,
 {\rm P}\!\exp\left\{
 \int_0^\infty \!d\tilde{\eta}
\left(
\begin{array}{cc}
h(\tilde{\eta}) & -m^2\\
-1 & 0
\end{array}
\right)
\right\}
\left(
\begin{array}{c}
\pi(\infty)\\
\phi(\infty)
\end{array}
\right)
\right]
\right)^2
\nonumber \\
&+\sum_{\small\begin{array}{l}
\{\pi(\infty),\phi(\infty)\}
\\{\rm negative }
\end{array}}
\left(
\tanh
\left[
(1 \; 0)\,
 {\rm P}\!\exp\left\{
 \int_0^\infty \!d\tilde{\eta}
\left(
\begin{array}{cc}
h(\tilde{\eta}) & -m^2\\
-1 & 0
\end{array}
\right)
\right\}
\left(
\begin{array}{c}
\pi(\infty)\\
\phi(\infty)
\end{array}
\right)
\right]
-1\right)^2.
\end{align}
The learning process is equivalent to the following gradient flow equation with a fictitious time variable $\tau$,
\begin{align}
\frac{\partial h(\eta,\tau)}{\partial \tau} =\frac{\partial E[h(\eta,\tau)]}{\partial h(\eta,\tau)} \, .
\end{align}

For the training of our numerical experiment using the experimental data,
we have chosen the initial configuration of $h(\eta)$ as a constant (which corresponds to a pure AdS metric). For 
a constant $h(\eta)=h$, the error function can be explicitly evaluated with
\begin{align}
\pi(0)=\frac{1}{\lambda_+-\lambda_-}
\left(
\lambda_+(\pi(\eta_{\rm ini})-\lambda_-\phi(\eta_{\rm ini}))e^{-\lambda_+\eta_{\rm ini}}
+\lambda_-(-\pi(\eta_{\rm ini})+\lambda_+\phi(\eta_{\rm ini}))e^{-\lambda_-\eta_{\rm ini}}
\right)
\end{align}
where $\lambda_\pm\equiv \frac12(-h\pm\sqrt{h^2+4m^2})$ is the eigenvalue of the matrix which is path-ordered.
Using this expression, we find that at the initial epoch of the training the function $h(\eta)$ is updated 
by an addition of a function of the form $\exp[(\lambda_+-\lambda_-)\eta]$
and of the form $\exp[-(\lambda_+-\lambda_-)\eta]$. This means that the update is effective in two regions:
near the black hole horizon $\eta \approx 0$ and near the AdS boundary $\eta \approx \infty$.

Normally in deep learning the update is effective near the output layer because any back propagation could be
suppressed by the factor of the activation function. However our example above shows that the update
near the input layer is also updated. The reason for this difference is that in the example above 
we assumed $\lambda=0$ to solve the error function explicitly, and it means that the activation function is trivial.
In our numerical simulations where $\lambda \neq 0$, the back propagation is expected to be suppressed near
the input layer.

\section{Black hole metric and coordinate systems}

Here we summarize the properties of the bulk metric and the coordinate frame which we prefer to use in the main text.

The 4-dimensional AdS Schwarzschild black hole metric is given by
\begin{align}
ds^2 = -f(r)dt^2+\frac{1}{f(r)} dr^2 + \frac{r^2}{L^2} \sum_{i=1}^{2}dx_i^2, \quad
f(r)\equiv \frac{r^2}{L^2}\left(1-\frac{r_0^3}{r^3}\right)
\label{metrica}
\end{align}
where $L$ is the AdS radius, and $r=r_0$ is the location of the black hole horizon. $r=\infty$
corresponds to the AdS boundary. To bring it to the form \eqref{genericm}, we make a coordinate transformation
\begin{align}
r=r_0 \left(\cosh \frac{3\eta}{2L}\right)^{2/3}.
\end{align}
With this coordinate $\eta$, the metric is given by
\begin{align}
ds^2 = -f(\eta)
dt^2+d\eta^2 + g(\eta)
\sum_{i=1}^{2}dx_i^2,
\quad
f(\eta)\equiv
\frac{r_0^2}{L^2}\left(\cosh \frac{3\eta}{2L}\right)^{-2/3}\!
\left(\sinh \frac{3\eta}{2L}\right)^{2}, \quad
g(\eta) \equiv
\frac{r_0^2}{L^2} \left(\cosh \frac{3\eta}{2L}\right)^{4/3}.
\end{align}
The AdS boundary is located at $\eta = \infty$ while the black hole horizon
resides at $\eta=0$. The function $h(\eta)$ appearing in the scalar field equation \eqref{sceq} is
\begin{align}
h(\eta)\equiv \partial_\eta \log \sqrt{f(\eta)g(\eta)^{d-1}}
= \frac{3}{L} \coth \frac{3\eta}{L} \, .
\label{hcoth}
\end{align}
The $r_0$ dependence, and hence the temperature dependence, disappears because our
scalar field equation \eqref{sceq} assumes time independence and $x_i$-independence.
This $h(\eta)$ is basically the invariant volume of the spacetime, and is important in the sense that
a certain tensor component of the vacuum Einstein equation
coming from
\begin{align}
S_{\rm E} = \int \! d^4x \sqrt{-\det g}\left(R + \frac{6}{L^2}\right)
\end{align}
results in a closed form
\begin{align}
-\frac{9}{L^2} + \partial_\eta h(\eta) + h(\eta)^2 = 0 \, .
\label{Eeq}
\end{align}
It can be shown that the ansatz \eqref{metrica} leads to a unique metric solution for the vacuum 
Einstein equations, and the solution is given by \eqref{hcoth} up to a constant shift of $\eta$.
Generically, whatever the temperature is, and whatever the matter energy momentum tensor is, 
the metric function $h(\eta)$ behaves as $h(\eta)\approx 1/\eta$
near the horizon $\eta \approx 0$, and goes to a constant (proportional to the AdS radius $L$) at the AdS boundary 
$\eta\approx\infty$. 

One may try to impose some physical condition on $h(\eta)$. In fact, the right hand side of \eqref{Eeq}
is a linear combination of the energy momentum tensor, and generally we 
expect that the energy momentum tensor is subject to various energy conditions, 
which may constrain the $\eta$-evolution of $h(\eta)$. Unfortunately
it turns out that a suitable energy condition for constraining $h(\eta)$ is not available, within our search.
So, non-monotonic functions in $\eta$ are allowed as a learned metric. 

\section{Details about our coding for the learning}

\subsection{Comments on the regularization}

Before getting into the detailed presentation of the coding, let us make some comments on the
effect of the regularization $E_{\rm reg}$ and the statistical analysis of the learning trials. 

First, we discuss the meaning of $E_{\rm reg}$ in \eqref{loss}. In the first numerical experiment 
for the reproduction of the AdS Schwarzschild black hole metric we took
\begin{align}
E_{\rm reg}^{(1)} \equiv 3\times 10^{-3}\sum_{n=1}^{N-1}
(\eta^{(n)})^4 \left(h(\eta^{(n+1)})-h(\eta^{(n)})\right)^2 
\propto  \int d\eta\, (h'(\eta)\eta^2)^2.
\end{align}
This regularization term works as a selection of the metrics which are smooth. We are interested in the
metric with which we can take a continuum limit, so a smooth $h(\eta)$ is better for our physical interpretation.
Without $E_{\rm reg}$, the learned metrics are
far from the AdS Schwarzschild metric: see Fig.\ref{fig:reg}  for an example of the learned metric
without $E_{\rm reg}$. Note that the example in Fig.~\ref{fig:reg}  achieves the accuracy which is 
the same order as that of the learned metric with $E_{\rm reg}$. So, in effect, this regularization term
does not spoil the learning process, but actually picks up the metrics which are smooth, among the
learned metrics achieving the same accuracy.

Second, we discuss how the learned metric shown in Fig.~\ref{fig:adslearn} is generic, for the case of the 
first numerical experiment.
We have collected results of 50 trials of the machine learning, and the statistical analysis is presented in Fig.~\ref{fig:adslearn} (c). It is shown that the metric in the asymptotic region is quite nicely learned, 
and we can conclude that the asymptotic AdS spacetime has been learned properly.
On the other hand, for the result in the region near the black hole horizon, the learned metric
reproduces qualitatively the behavior around the horizon, but quantitatively it deviates from the true metric.
This could be due to the discretization of the spacetime.

Third, let us discuss the regularization for the second numerical experiment for the emergence of the
metric for the condensed mater material data. The regularization used is
\begin{align}
E_{\rm reg}&= E_{\rm reg}^{(1)} + E_{\rm reg}^{(2)}
\nonumber \\
&= 3\times 10^{-3}\sum_{n=1}^{N-1}(\eta^{(n)})^4 \left(h(\eta^{(n+1)})-h(\eta^{(n)})\right)^2 
\; + 
\; 
c_{\rm reg}^{(2)} \left(h(\eta^{(N)})-1/\eta^{(N)}\right)^2,
\end{align}
with $c_{\rm reg}^{(2)}=10^{-4}$. The second term is to fit the metric $h(\eta)$ 
near the horizon to the value $1/\eta$, because $1/\eta$ behavior is expected for any regular horizons.
In Fig.~\ref{fig:stat}, we present our statistical analyses of the obtained metrics for two other distinct choices
of the regularization parameter: $c_{\rm reg}^{(2)}=0$ and $c_{\rm reg}^{(2)}=0.1$.
For $c_{\rm reg}^{(2)}=0$, there is no regularization $E_{\rm reg}$, so the metric goes down to a negative 
number at the horizon. For $c_{\rm reg}^{(2)}=0$, which is a strong regularization, the metric is
almost completely fixed to a value $1/\eta$ with $\eta = \eta^{(N)}$.
For all cases, the learned metrics achieve a loss $\approx 0.02$, so the system is successfully learned.
The only difference is how we pick up "physically sensible" metrics among many learned metrics.
In Fig.~\ref{fig:expres}, 
we chose $c_{\rm reg}^{(2)}=10^{-4}$ which is in between the values used in Fig.~\ref{fig:stat},
because the deviation of the metric near the horizon is of the same order as that near the asymptotic region.

\begin{figure}[t]
\includegraphics[width=8cm]{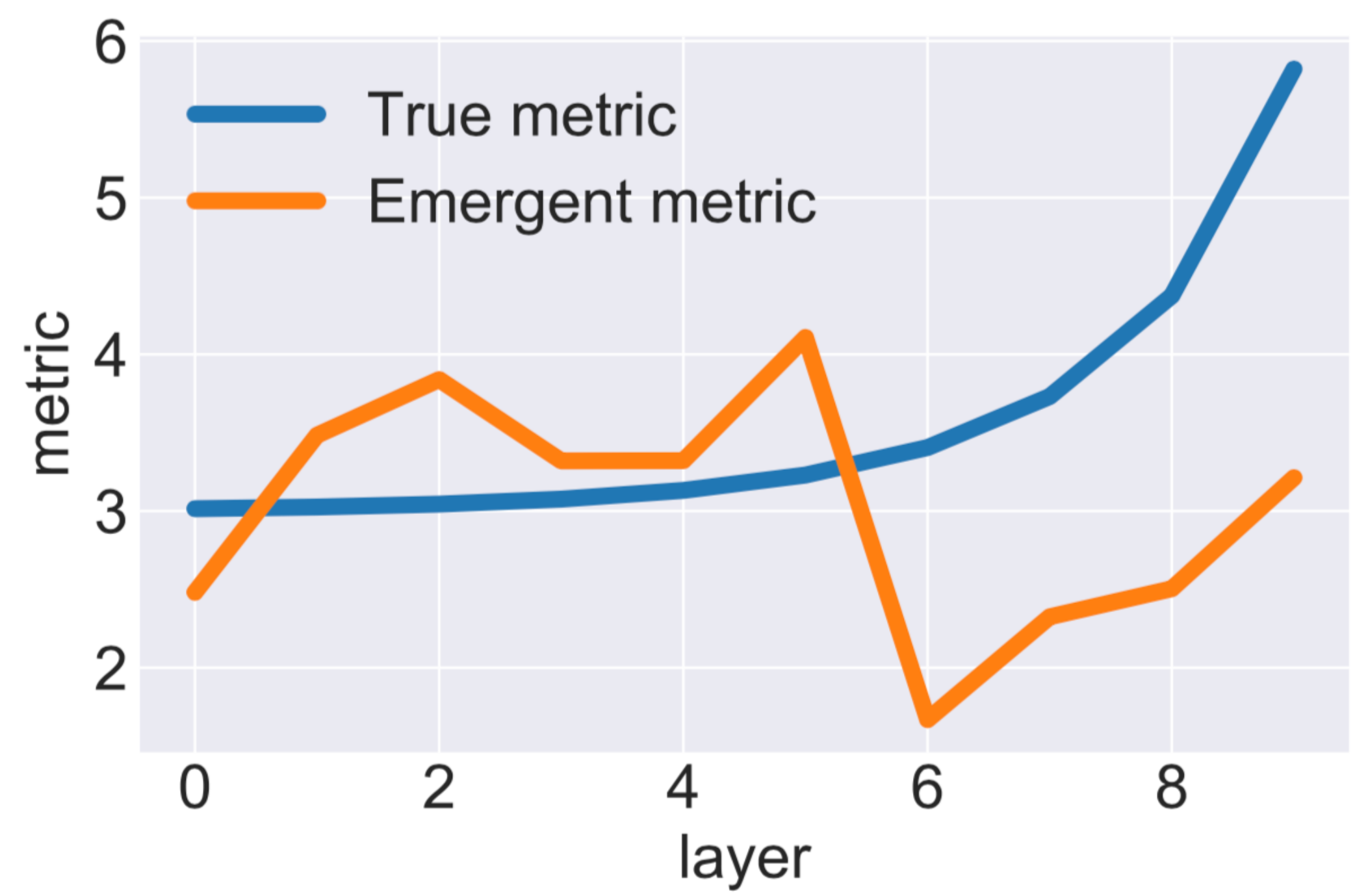}
\caption{
A learned metric with a high accuracy, without the use of the regularization $E_{\rm reg}$. 
The used setup is the same as what we used for the reproduction of the AdS Schwarzschild metric. }
\label{fig:reg}
\end{figure}

\begin{figure}[t]
\includegraphics[width=16cm]{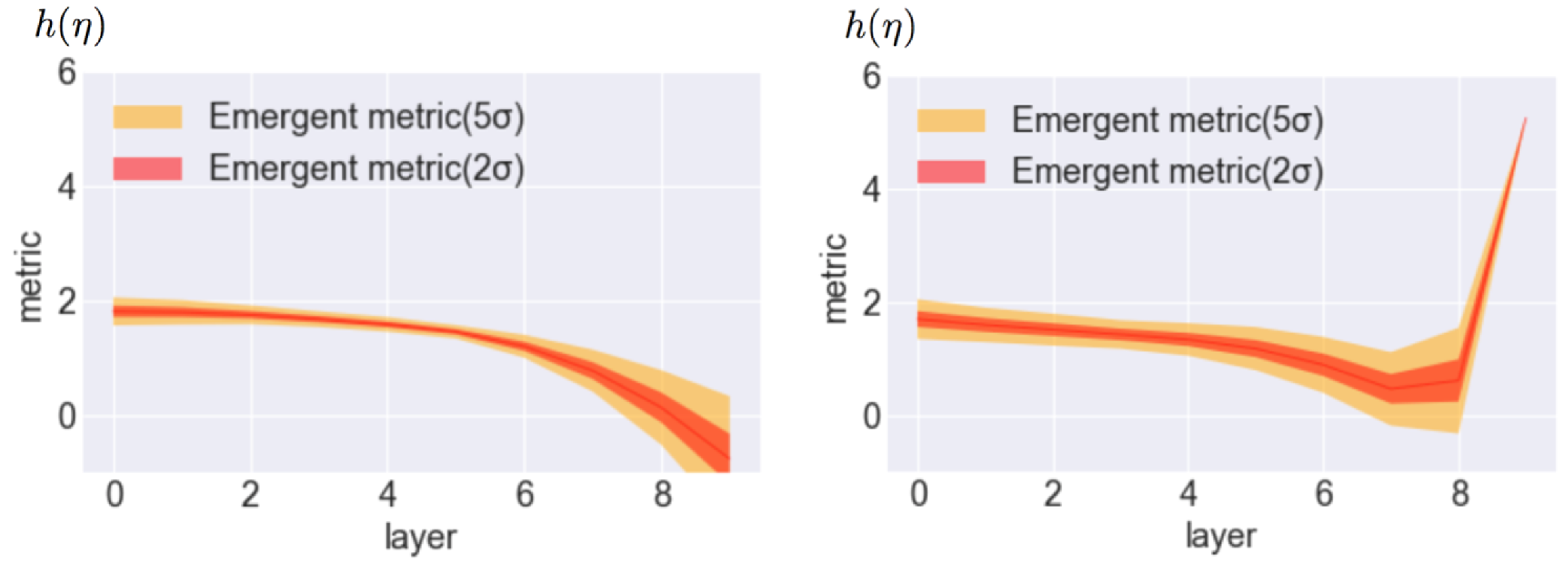}
\caption{Statistical results of the obtained 13 metrics. Left: $c_{\rm reg}^{(2)}=0$. Right: $c_{\rm reg}^{(2)}=0.1$. }
\label{fig:stat}
\end{figure}

\subsection{Numerical experiment 1: Reconstructing AdS Schwarzschild black hole}

We have performed two independent numerical experiments: The first one is about the reconstruction of the
AdS Schwarzschild black hole metric, and the second one is about the emergence of a metric from 
the experimental data of a condensed matter material. Here we explain details about the coding and the setup,
for each numerical experiment.

In the first numerical experiment, we fix the mass of the scalar field $m^2$ and coupling constant in potential $V(\phi) = \frac{\lambda}{4} \phi^4$ to
\begin{align}
m^2 = -1,
\quad
\lambda = 1,
\end{align}
and prepare data $\{ (\bar{x}^{(1)} , \bar{y}) \}$ to train the neural network.
The training data is just a list of initial pairs of $\bar{x}^{(1)} = (\phi, \pi)$ and corresponding answer signal $\bar{y}$.
We regard $\bar{x}^{(1)} = (\phi, \pi)$ as field values at the AdS boundary, and define the answer signal so that it represents whether they are permissible or not when they propagate toward the black hole horizon.
More explicitly, what we do is the iteration defined below:
\begin{enumerate}
\item{randomly choose $\phi \in [0 , 1.5 ]$, $\pi \in [-0.2, 0.2]$ and regard them as input : $\bar{x}^{(1)} = 
\begin{pmatrix}
\phi \\
\pi
\end{pmatrix}
$.}
\item{propagate it by E.O.M \eqref{prop} with AdS Schwarzschild metric \eqref{AdSS} from 
$
\begin{pmatrix}
\phi(\eta_\text{ini}) = \phi \\
\pi(\eta_\text{ini}) = \pi
\end{pmatrix}
$ to 
$
\begin{pmatrix}
\phi(\eta_\text{fin}) \\
\pi(\eta_\text{fin})
\end{pmatrix}
$.}
\item{calculate consistency $F$, i.e. right hand side of \eqref{Ff}, and define the answer signal :
$
\bar{y} =
\left\{ \begin{array}{ll}
0 & \text{if $F < 0.1$} \\
1 & \text{if $F > 0.1$}\\
\end{array} \right.$
.
}
\end{enumerate}
To train the network appropriately, it is better to prepare a data containing roughly equal number of $\bar{y}=0$ samples and $\bar{y}=1$ samples.
We take a naive strategy here:
If the result of step 3 becomes $\bar{y} = 0$, we add the sample $(\bar{x}^{(1)}, \bar{y})$ to the positive data category, if not, we add the sample to the negative data category.
Once the number of samples of one category saturates to $10^3$, we focus on collecting samples in another category.
After collecting both data, we concatenate positive data and negative data and regard it as the total data for the training:
\begin{align}
\text{Training data }
D
&=
\Big(
\text{$10^3$ positive data}
\Big)
\oplus
\Big(
\text{$10^3$ negative data}
\Big),
\quad
\text{where}
\left\{ \begin{array}{ll}
\text{positive data} = \{ (\bar{x}^{(1)}, \bar{y}=0) \}  \\
\text{negatve data} = \{ (\bar{x}^{(1)}, \bar{y}=1) \} 
\end{array} \right.
.
\notag
\end{align}
Besides it, we prepare the neural network \eqref{nns} with the restricted weight \eqref{W}.
The only trainable parameters are $h(\eta^{(n)})$, and the purpose of this experiment is to see whether trained $h(\eta^{(n)})$ are in agreement with AdS Schwarzschild metric \eqref{AdSS} encoded in the training data implicitly.
To compare $\bar{y}$ and neural net output $y$, we make following final layer.
First, we calculate $F \equiv \pi(\eta_\text{fin})$ (which is the r.h.s.~of \eqref{Ff} in the limit $\eta_\text{fin}\to 0$), and second, we define $y \equiv t(F)$ where
\begin{align}
t(F)
=
\Big[
\tanh\Big(100 (F - 0.1) \Big)
-
\tanh\Big(100 (F + 0.1) \Big)
+2
\Big]/2
.
\label{t(F)}
\end{align}
We plot the shape of $t(F)$ in Figure \ref{fig:t_function}.
Before running the training iteration, we should take certain initial values for $h(\eta^{(n)})$. We use the initial  $h(\eta^{(n)}) \sim \mathcal{N} (1/\eta^{(n)}, 1 ) $ (which is a gaussian distribution), because any black hole horizon is characterized by 
the $1/\eta^{(n)}$ behavior at $\eta^{(n)}\approx 0$.
\footnote{Note that we do not teach the value of $h(\eta)$ at the AdS boundary, i.e. $3$ in our case.}
After setting the initial values for the trained parameters, we repeat the training iteration:
\begin{enumerate}
\item{randomly divide the training data to direct sum : $D =(\text{mini data 1}) \oplus (\text{mini data 2}) \oplus \dots \oplus (\text{mini data 200})  $.}
\item{calculate loss \eqref{loss} and update $h(\eta^{(n)})$ by Adam optimizer \cite{kingma2014adam} for each mini data.}
\end{enumerate}
When the target loss function \eqref{loss} becomes less than 0.0002, we stop the iteration 1 and 2.

\subsection{Numerical experiment 2: Emergent metric from experimental data}
\begin{figure}[t]
\begin{minipage}{0.45\hsize}
	\includegraphics[width=8cm]{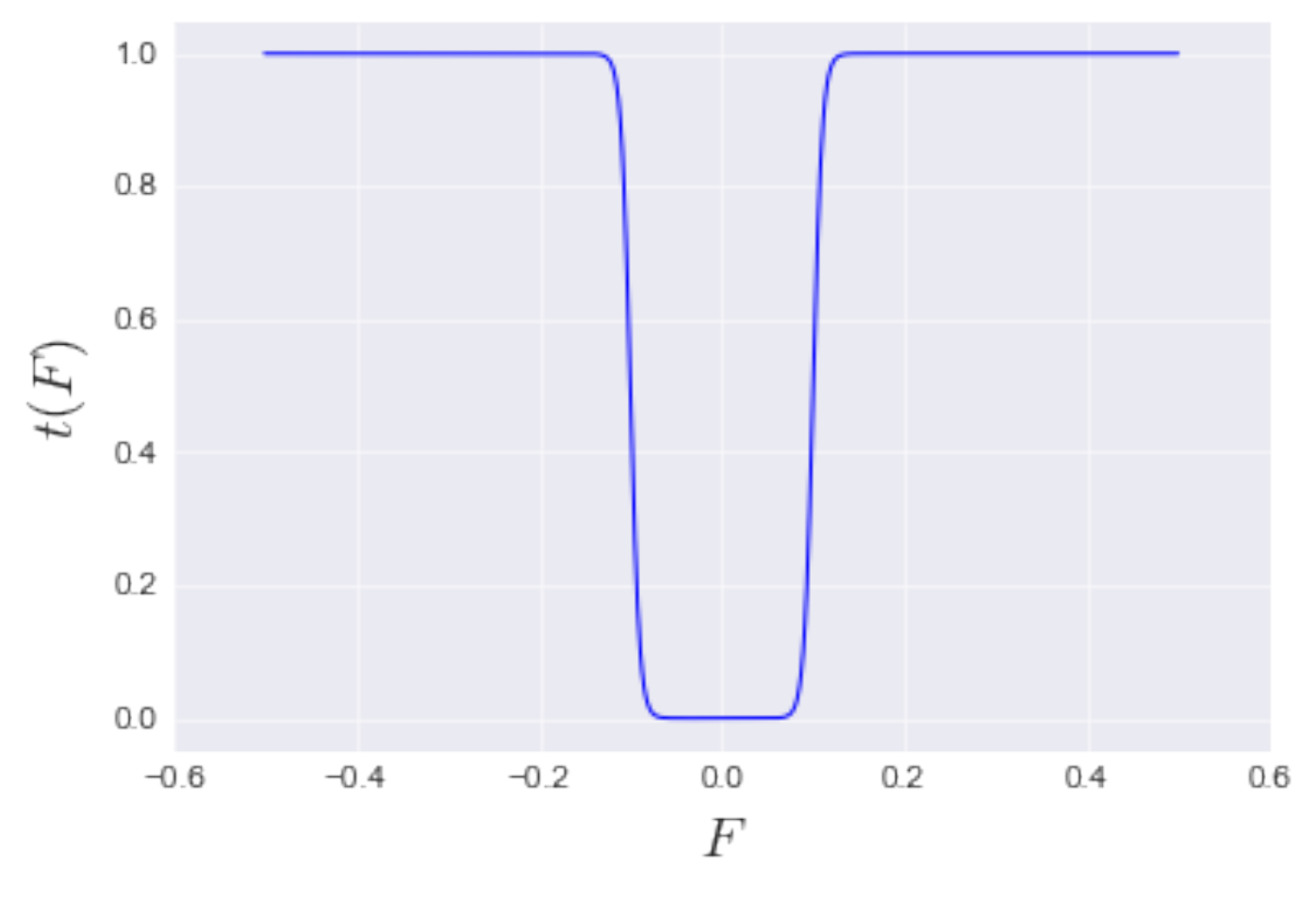}
	\caption{
		Final layer function $t(F)$ in \eqref{t(F)}.
	}
	\label{fig:t_function}
\end{minipage}
\hspace{.5cm}
\begin{minipage}{0.45\hsize}
	\includegraphics[width=8cm]{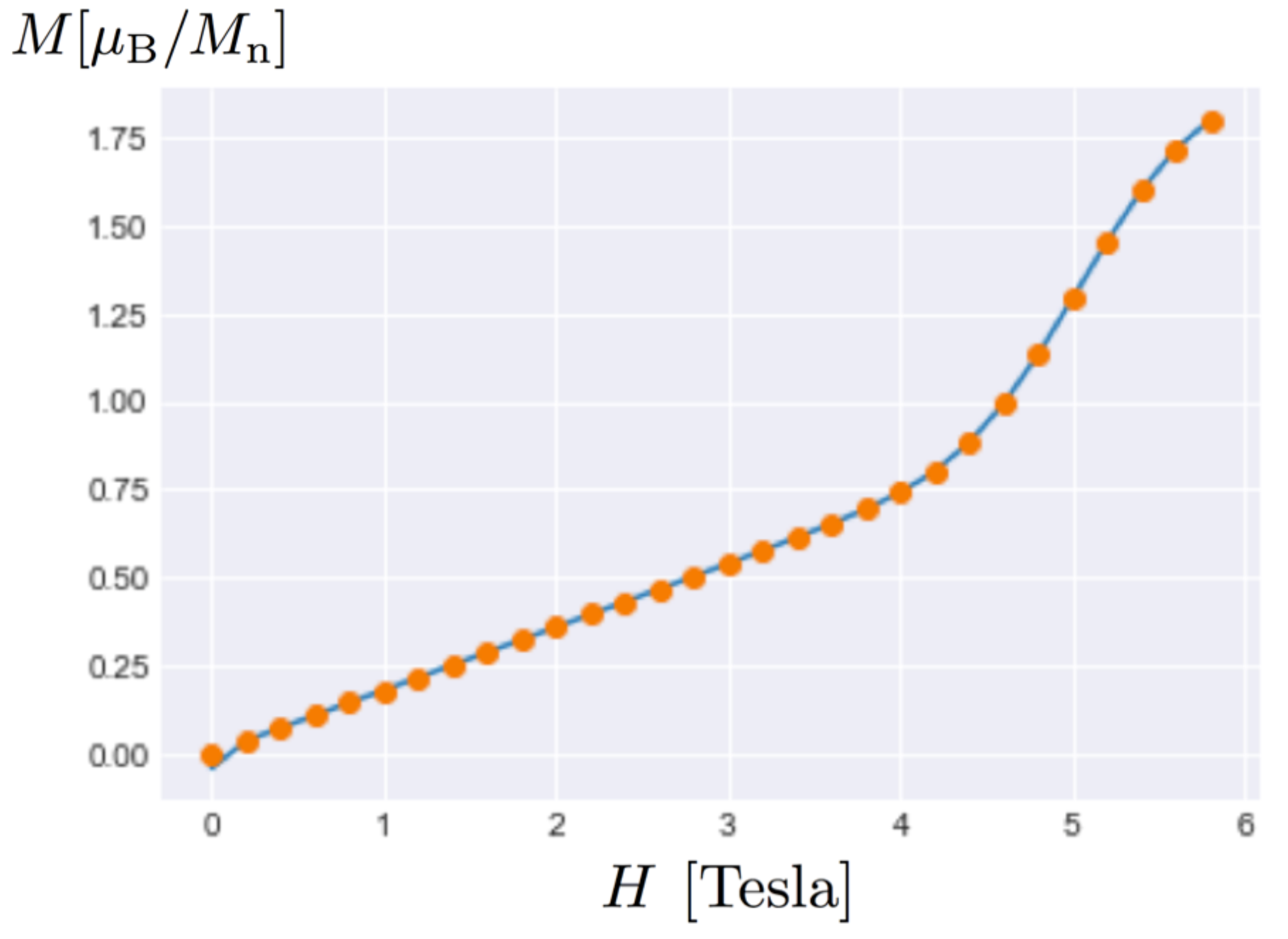}
	\caption{
		Experimental data of magnetization $(M)$ versus magnetic field $(H)$ and its polynomial fitting.
	}
	\label{fig:exp}
\end{minipage}
\end{figure}

As a next step, we perform the second numerical experiment.
In this case, we use experimental data \cite{sakai} composed by pairs of strength of magnetic field $H$ and corresponding magnetic response $M$ of Sm$_{0.6}$Sr$_{0.4}$MnO$_3$ at the temperature 155K. 
To pad the data, we plot the experimental paired $(H, M)$ values to a 2-dimensional scatter plot and fit it by using a polynomial with respect to $H$ up to 15-th order (see Fig.~\ref{fig:exp}), and call it $f(H)$.
By using this $f(H)$, we prepare the training data $\{ (\bar{X}^{(1)}, \bar{y} ) \}$ as follows:
\begin{enumerate}
\item{randomly choose $H \in [0, 6], M \in [0, 2]$ and regard them as input : 
$
\bar{X}^{(1)} = 
\begin{pmatrix}
H \\
M
\end{pmatrix}
$
}
\item{define the answer signal : 
$
\bar{y}
=
\left\{ \begin{array}{ll}
0 & \text{if } M \in [f(H) - \text{noise},  f(H)+\text{noise}] \\
1 & \text{otherwise}\\
\end{array} \right.
$
where the noise $\sim \mathcal{N}(0, 0.1)$
}
\end{enumerate}
We prepare $10^4$ positive data and $10^4$ negative data as same as done in the first numerical experiment.
See Fig.~\ref{fig:sakai} for a padding of the obtained data.
On the neural network, we insert an additional layer as the 1st layer \eqref{nnpia2}.
In addition to the values for $h(\eta^{(n)})$, we update $\alpha, \beta$ in \eqref{nnpia2} and $m^2, \lambda$ in \eqref{prop} and \eqref{W} with $V(\phi) = \frac{\lambda}{4} \phi^4$.
As one can notice, there is $m^2$ in the definitions for $\Delta_\pm$, so \eqref{nnpia2} includes $m^2$ implicitly.
The training is performed in the same manner as the first numerical experiment.
We use 10-layered neural network in our numerical experiments.
When the target loss function \eqref{loss} goes smaller than 0.02, we stop the learning.
Initial conditions for the network are taken as $h(\eta^{(n)}) \sim \mathcal{N}(2, 1), m^2 \sim \mathcal{N}(2, 1), \lambda \sim \mathcal{N}(1,1)$ and $\alpha, \beta \sim [-1, 1]$.


\section{Comments on the conformal dimensions}

Here we review the critical exponents for a magnetic system, which are described by a scalar field near the critical point. 
On $D$-dimensional space $(D=d-1)$, the correlation function of the scalar field behaves as
\begin{align}
G(x)\sim |x|^{-(D-2+\eta)} 
\end{align}
at the critical temperature, where $\eta$ is the anomalous dimension. Thus, the scaling dimension of the scalar is given by 
\begin{align}
\Delta=\frac{D-2+\eta}{2}.
\end{align}

The critical exponent $\delta$ is defined as 
\begin{align}
M\sim H^{1/\delta}
\end{align}
at the critical temperature, 
\textit{i.e.}, $\delta$ characterizes how the magnetization $M$ depends on the magnetic field $H$ near $H = 0$. 
It is known (see \textit{e.g.} \cite{DiFrancesco:1997nk}) that the scaling hypothesis relates the critical exponents $\delta$ and $\eta$ as 
\begin{align}
\delta=\frac{D+2-\eta}{D-2+\eta}\,.
\label{scaling_law}
\end{align}

The critical exponent $\delta$ should be positive because the magnetization $M$ should vanish when the magnetic field $H$ is turned off. Thus, the scaling law \eqref{scaling_law} implies that the anomalous dimension $\eta$ satisfies $\eta<D+2$. Therefore, the scaling dimension $\Delta$ should be bounded as $\Delta<D$. In particular, setting $D=3$,  we should have $\Delta<3$. 

However, in our numerical experiment using the magnetic response data of the material Sm$_{0.6}$Sr$_{0.4}$MnO$_3$ at 155 K, from the obtained data we
can calculate the  conformal dimension, $\Delta_+ = 4.89 \pm 0.32$.
The estimated value of the conformal dimension 
is larger than the bound $\Delta_+<3$, and we have to be careful in
the interpretation of the value here. 

Let us discuss several possible reasons for the violation of the bound. 
In fact, we use a scalar model
which does not properly reflect the spin structure of the operator.
For holographic treatment of the magnetization, several ways were proposed: see
\cite{Iqbal:2010eh,Hashimoto:2013bna,Cai:2014oca,Cai:2015jta,Yokoi:2015qba}. 
Depending on the models, the identification of the conformal dimension could be different.

Another reason is that when we compute $\Delta_+$ numerically, we set $\eta_{\rm ini}=1$ to reduce the computational task. If we chose $\eta_{\rm ini}$ to take a much larger value $\eta_{\rm ini}/L \gg 1$, the
extent of the violation would have been milder.

We also speculate that the temperature 155K we chose for the analyses may not be close enough to the critical temperature. 
In addition, because the order of the phase transition is not evident in the experimental data, the scaling law discussed above may not be applied.
Of course, even if the temperature is near the critical temperature, there is no persuasive reason that the material Sm$_{0.6}$Sr$_{0.4}$MnO$_3$ can be described holographically by a classical bulk scalar field. The simulation is just a demonstration of how our DL is used for the given experimental data, and we do not take the violation of the bound as a serious problem in this letter. It is more interesting to find a material such that the scaling dimension computed from our DL agrees with the critical exponents estimated from the experimental data. If we have such a material, the agreement suggests that it has a holographic dual. 


\vspace{5mm}

\end{widetext}

\end{document}